\documentclass[journal,transmag]{IEEEtran}

\usepackage[utf8]{inputenc}
\usepackage[T1]{fontenc}
\usepackage{subfiles}
\usepackage[english]{babel} 
\usepackage{framed}
\usepackage{amsmath}

\usepackage{amssymb}
\usepackage{listings}
\usepackage{graphicx}
\usepackage{afterpage}
\usepackage{mathtools}
\usepackage{xspace}
\usepackage{siunitx}
\usepackage{subcaption}
\usepackage{hyperref}

\newcommand{\ez}{\ensuremath{ \mathbf{e}_z}\xspace}
\newcommand{\Cmat}{\ensuremath{ \mathbf{C} }\xspace}
\newcommand{\Dmat}{\ensuremath{ \mathbf{D} }\xspace}
\newcommand{\Kmat}{\ensuremath{ \mathbf{K} }\xspace}
\newcommand{\mtab}[2]{\ensuremath{ \left[\begin{array}{cc}{#1}&{#2}\end{array}\right] }}
\newcommand{\mtba}[2]{\ensuremath{ \left[\begin{array}{c}{#1}\\ {#2}\end{array}\right] }}
\newcommand{\mtbb}[4]{\ensuremath{ \left[\begin{array}{cc}{#1}&{#2}\\ {#3}&{#4}\end{array}\right] }}
\newcommand{\zA}{\ensuremath{\mathbf{A}}\xspace}
\newcommand{\zB}{\ensuremath{\mathbf{B}}\xspace}
\newcommand{\zH}{\ensuremath{\mathbf{H}}\xspace}
\newcommand{\zJ}{\ensuremath{\mathbf{J}}\xspace}
\newcommand{\zr}{\ensuremath{\mathbf{r}}\xspace}

\newcommand{\pd}[2]{\frac{\partial{#1}}{\partial{#2}}}
\newcommand{\ud}[1]{\,\mbox{d}{#1}}

\newcommand{\pdx}[1]{\pd{#1}{{x}}}
\newcommand{\pdy}[1]{\pd{#1}{{y}}}
\newcommand{\udOmegaprime}{\ud{{\Omega^\prime}}}

\catcode`@=11
\newcommand{\frownfill}{\ensuremath{\scriptscriptstyle\m@th\mathord\frown}}
\newcommand{\bow}[1]{\ensuremath{\vbox{\m@th\ialign{##\crcr
        \hfil\frownfill\hfil\crcr\noalign{\kern-0.2\p@\nointerlineskip}
        $\hfil\displaystyle{#1}\hfil$\crcr}}}}
\newcommand{\bbow}[1]{\ensuremath{\vbox{\m@th\ialign{##\crcr
        \hfil\frownfill\hfil\crcr\noalign{\kern-0.7\p@\nointerlineskip}
        \hfil\frownfill\hfil\crcr\noalign{\kern-0.3\p@\nointerlineskip}
        $\hfil\displaystyle{#1}\hfil$\crcr}}}}
\newcommand{\volt}[1]{\protect\bow{#1}}
\newcommand{\flux}[1]{\protect\bbow{#1}}
\newcommand{\afit}{\ensuremath{\volt{\mathrm{\mathbf{a}}}}\xspace}
\newcommand{\jfit} {\ensuremath{\flux{\mathrm{\mathbf{j}}}}\xspace}
\newcommand{\Knu}{\ensuremath{\Kmat_{\nu}}}

\newcommand{\qcm}{\;,}
\newcommand{\qpt}{\;.}
\newcommand{\qscn}{\;;\\}
\newcommand{\idx}[1]{_{\text{#1}}}
\newcommand{\idxi}[2]{_{{\text{#1}},#2}}
\newcommand{\idxx}[2]{_{\text{#1,#2}}}
\newcommand{\diag}[1]{\text{diag}\{#1\}}      
\newcommand{\up}[1]{^{(#1)}}
\newcommand{\zw}{\mathbf{w}}
\usepackage{bm}
\newcommand{\etabf}{\bm{\eta}}
\newcommand{\mubf}{\bm{\mu}}
\newcommand{\nubf}{\bm{\nu}}
\newcommand{\Knutld}{\Kmat_{\tilde{\nubf}}}

\begin{document}
\title{Magnetic Field Simulation with Data-Driven Material Modeling}

\author{\IEEEauthorblockN{Herbert De~Gersem\IEEEauthorrefmark{1}\IEEEauthorrefmark{2}\IEEEauthorrefmark{3},
Armin Galetzka\IEEEauthorrefmark{1}\IEEEauthorrefmark{2},
Ion Gabriel Ion\IEEEauthorrefmark{1}\IEEEauthorrefmark{2}, 
Dimitrios Loukrezis\IEEEauthorrefmark{1}\IEEEauthorrefmark{2}, and
Ulrich Römer\IEEEauthorrefmark{4}}
\IEEEauthorblockA{\IEEEauthorrefmark{1}Institut für Teilchenbeschleunigung und Elektromagnetische Felder, TU Darmstadt, Germany}
\IEEEauthorblockA{\IEEEauthorrefmark{2}Graduate School Computational Engineering, TU Darmstadt, Germany}
\IEEEauthorblockA{\IEEEauthorrefmark{3}Wave Propagation and Signal Processing Research Group, KU Leuven - Kulak, Belgium}
\IEEEauthorblockA{\IEEEauthorrefmark{4}Institut für Dynamik und Schwingungen, TU Braunschweig, Germany}
\thanks{Manuscript received October 1, 2019; revised MMM DD, YYYY. 
Corresponding author: H. De~Gersem (email: degersem@temf.tu-darmstadt.de).}}

\markboth{Accepted for publication in IEEE Transactions on Magnetics, 8th June 2020, DOI: \href{https://doi.org/10.1109/TMAG.2020.3002092}{10.1109/TMAG.2020.3002092}}%
{De~Gersem \MakeLowercase{\textit{et al.}}: Magnetic Field Simulation with Data-driven Material Modeling}

\IEEEtitleabstractindextext{%
\begin{abstract}
This paper develops a data-driven magnetostatic finite-element (FE) solver which directly exploits measured material data instead of a material curve constructed from it. The distances between the field solution and the measurement points are minimized while enforcing Maxwell's equations. The minimization problem is solved by employing the Lagrange multiplier approach. The procedure wraps the FE method within an outer data-driven iteration. The method is capable of considering anisotropic materials and is adapted to deal with models featuring a combination of exact material knowledge and measured material data. Thereto, three approaches with an increasing level of intrusivity according to the FE formulation are proposed. The numerical results for a quadrupole-magnet model show that data-driven field simulation is feasible and affordable and overcomes the need of modeling the material law.
\end{abstract}

}

\maketitle
\IEEEpeerreviewmaketitle

\section{Introduction}

\IEEEPARstart{E}{lectromagnetic} field problems are governed by the Maxwell equations and by constitutive laws describing the material behavior. The former are exact, whereas the latter may come together with a substantial level of uncertainty, especially for exotic materials. No matter how fine the spatial and temporal scales are resolved, the accuracy of the overall field simulation may heavily depend on the provided material data. Only in rare cases, ab-initio material models are available, not to mention the exuberant computational effort they require, hampering their application within field models \cite{Daniel_2008aa,Vanoost_2016aa}. More often, sets of material data are given. The standard treatment assumes the essential material behavior to be known by means of a parametrized material curve, e.g., as represented by an analytical expression or by a spline approximation, for which the parameters are found by regression \cite{Pechstein_2006aa}, machine learning \cite{Veeramani_2009aa} or any other appropriate fitting technique. Due to measurement errors and ambient influences, the obtained data can be considered as uncertain. Consequently, these uncertain data propagates through the field model and renders the model output uncertain as well \cite{Bartel_2013ab}. When the material behavior is itself stochastic, in particular, if it is represented with a random field over the computational domain, a stochastic model based on a Karhunen-Lo\`eve expansion is due \cite{Jankoski_2017aa,Jankoski_2019aa}.

For many non-trivial materials, the choice of an appropriate material model is not easy. Also the regression algorithm to determine the material-model parameters is still subject of active research, especially when one wants to keep track of the uncertainties related to a finite measurement precision. When embedded in a field-simulation procedure, a material model needs to ensure a certain form of compatibility, e.g., numerical differentiability or monoticity \cite{Pechstein_2006aa}. Sometimes, part of the accuracy of the material model is sacrificed in favor of a higher compability with the field model \cite{Henrotte_2006aa}. Aligned with contemporary successes in the field of data science \cite{Baesens_2014aa}, material scientists consider the possibility to abandon  physically motivated material models altogether and, instead, deduce material behavior fully from larger data sets \cite{Rajan_2005aa,Gupta_2015aa}. Still, this does not change the field-simulation procedure itself, as it only replaces the embedded material model.

This paper goes one step further. Measured material data is no longer translated into a material model before its insertion in the field simulation. Instead, both simulation steps are carried out together by means of a so-called \emph{data-driven field solver}. In this paper, such a solver is constructed for the magnetostatic case and illustrated for a 2D quadrupole-magnet model. An extension of the idea to other electromagnetic formulations and to 3D simulation is, however, straightforward. The development closely follows the lines of \cite{Kirchdoerfer_2016aa}, where a data-driven field solver for mechanics is set up. This paper extends the work of \cite{Kirchdoerfer_2016aa} and describes how to account for a heterogeneous model in which exact and data-driven material behavior coexists.

The paper is structured as follows. In Section~II, preliminaries and notations for a finite-element (FE) field solver are given. In Section~III, a magnetostatic data-driven FE solver is derived for the case of a homogeneous material. In Section~IV, the data-driven FE solver is refurbished to allow for heterogeneous models, i.e., combining measured material data and exact material data for different model parts. Three approaches related to a different level of intrusiveness are formulated. Section~V illustrates the performance of each of the three approaches for the example of a quadrupole magnet.

\section{Magnetic Field Model}

\begin{figure}[t]
  \centering
  \includegraphics[width=8.5cm]{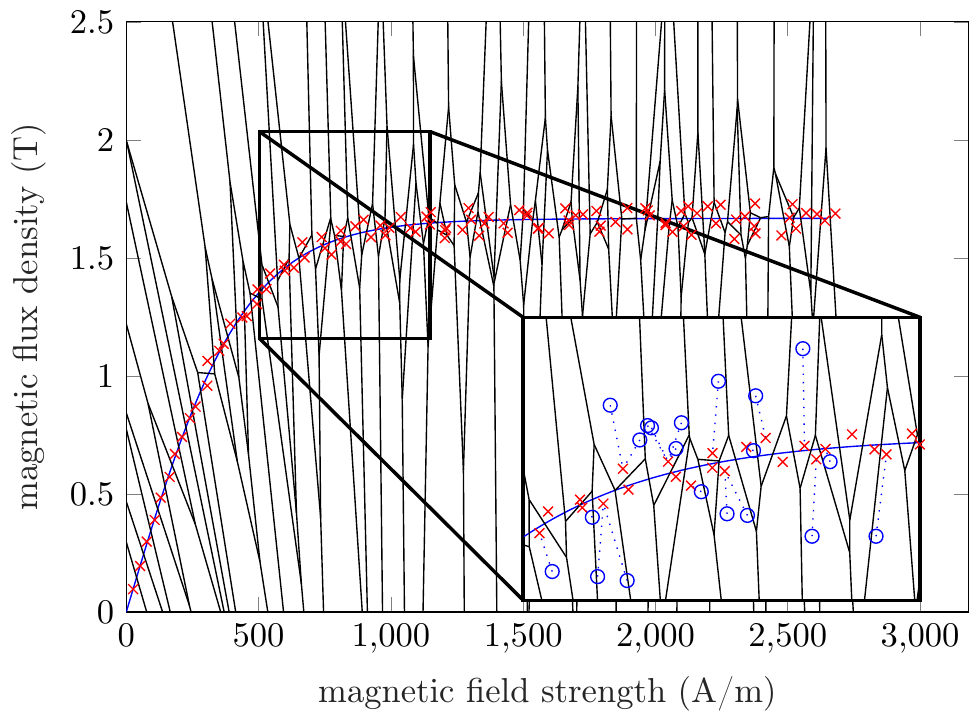}
  \caption{Measurement data (red crosses); Approximated material curve (blue line); Voronoi diagram corresponding to the minimization problem \eqref{eq:fe} (black mesh); detail of an intermediate field solution (blue circles) and their correspondence to the measurement data (blue dotted lines).}
  \label{fig:measurementdata}
\end{figure}

A magnetostatic field is described by Amp\`ere's and Gauss's laws:
\begin{subequations}
\begin{align}
  \label{eq:ampere} \nabla\times\zH &=\zJ \qscn
  \label{eq:gauss}  \nabla\cdot\zB  &=0 \qcm
\end{align}
\end{subequations}
where $\zH(\zr)$ is the magnetic field strength, $\zB(\zr)$ the magnetic flux density, $\zJ(\zr)$ the current density, and $\zr$ the spatial coordinate. For simple materials, the relation between $\zB$ and $\zH$ is given empirically by $\zB=\mubf\zH$ or, equivalently, $\zH=\nubf\zB$, where $\mubf(\zr)$ is the permeability and $\nubf(\zr)=\mubf^{-1}(\zr)$ is the reluctivity, which both can be tensorial. Here, in parts of the domain, materials are considered for which the relation between $\zB$ and $\zH$ is only known by, possibly noisy, measurement data (Fig.~\ref{fig:measurementdata}). This paper explicitly addresses the anisotropic case, in which the main axes of the material are assumed to coincide with the Cartesian coordinate system.

In a standard field simulation with an explicitly known reluctivity distribution $\nubf(\zr)$, one chooses to resolve \eqref{eq:gauss} by introducing the magnetic vector potential $\zA(\zr)$, such that $\zB=\nabla\times\zA$, and combining Maxwell's equations and the material model into the magnetostatic formulation
\begin{align}\label{eq:formulation}
  \nabla\times\left(\nubf\nabla\times\zA\right) &=\zJ \qcm
\end{align}
supplemented with appropriate boundary conditions. The Ritz-Galerkin FE procedure considers the weak form of \eqref{eq:formulation} according to a set of vectorial shape functions $\zw_i(\zr)$ and discretizes $\zA$ using the same set of functions:
\begin{subequations}
\begin{align}
  \label{eq:weakformulation}
  \int_\Omega \nubf\nabla\times\zA\cdot\nabla\times\zw_i \udOmegaprime
  &=\int_\Omega \zJ\cdot\zw_i \udOmegaprime \qscn
  \zA &=\sum_{j=1}^{N\idx{DoF}} \afit_j \zw_j \qcm
\end{align}
\end{subequations}
where $\afit$ contains the degrees of freedom (DoFs), $N\idx{DoF}$ is the number of DoFs and $\Omega$ is the computational domain. From here on, the development proceeds for the Cartesian 2D case. However, the message of this paper is equivalent for, e.g., the axisymmetric 2D case or the 3D case. Lowest-order test and trial functions $\zw_j(\zr)=N_j(x,y)\ez/\ell_z$, with $N_j(x,y)$ standard nodal hat functions defined on a triangulation of the 2D cross section, $\ez$ the unit vector in $z$-direction and $\ell_z$ the model's length in $z$-direction, are employed. The resulting system of equations reads
\begin{align}\label{eq:magnstat}
  \Knu\afit &=\jfit \qcm
\end{align}
where the stiffness matrix can be factorized as
\begin{align}\label{eq:Knu}
  \Knu &=\Cmat^T\Dmat_\Omega\Dmat_{\nubf}\Cmat \qpt
\end{align}
Here, $\Cmat$ denotes the algebraic representative of the curl operator, whereas $\Dmat_\Omega$ and $\Dmat_{\nubf}$ are diagonal matrices multiplying by the element volumes and expressing the material relation, i.e.,
\begin{subequations}
\begin{align}
  \jfit_i &=\int_\Omega \zJ\cdot\zw_i \udOmegaprime \qscn
  \Cmat_{xej} &=  \frac{1}{\ell_z}\pdy{N_i\up{e}} \qscn
  \Cmat_{yej} &= -\frac{1}{\ell_z}\pdx{N_i\up{e}} \qscn
  \Dmat_{\Omega,see} &=S_e\ell_z \qscn
  \label{eq:Dnu}\Dmat_{\nu,see} &=\nubf(\zr_e) \qcm
\end{align}
\end{subequations}
where $s\in\{x,y\}$, the superscript $\cdot\up{e}$ indicates a restriction to the triangle~$\mathcal{S}_e$ of the mesh, $e\in\mathcal{N}\idx{elem}$ with $\mathcal{N}\idx{elem}=\{1,\ldots N\idx{elem}\}$ and $N\idx{elem}$ the number of triangles, $S_e$ is the surface of $\mathcal{S}_e$, and $\zr_e$ is its center point. In a standard FE solver, $\Knu$ is assembled in one move. Here, however, access to the field data at the so-called \emph{material points} $\zr_e$ is needed. For lowest-order FEs as considered here, one material point per element is considered. For higher-order FEs, the material points should coincide with the quadrature points of the FE procedure.

\section{Data-driven Field Solver:\\Homogeneous Material}

At first, a data-driven field solver for a model with a \emph{homogeneous} material distribution is developed. The material behavior is described by two data sets $\mathcal{H}^\times_x=\{(H^\times_{p,x},B^\times_{p,x})\}$ and $\mathcal{H}^\times_y=\{(H^\times_{q,y},B^\times_{q,y})\}$ according to measurements along the $x$- and $y$-direction, respectively (Fig.~\ref{fig:measurementdata}). The set $\mathcal{H}^\times$ contains all possible combinations of any two data points of $\mathcal{H}^\times_x$ and $\mathcal{H}^\times_y$ and is formally defined by
\begin{align}
  \mathcal{H}^\times=\{(\zH^\times_{pq},\zB^\times_{pq})|(H^\times_{p,x},B^\times_{p,x})\in\mathcal{H}^\times_x,(H^\times_{q,y},B^\times_{q,y})\in\mathcal{H}^\times_y\} \qpt
\end{align}
The points $(\zH^\times_{pq},\zB^\times_{pq})$ live in a four-dimensional so-called \emph{phase space} \cite{Kirchdoerfer_2016aa}.

The magnetic fields at the material point $\zr_e$ are denoted by $(\zH^\circ_e,\zB^\circ_e)$ and are depicted by blue circles in Fig.~\ref{fig:measurementdata}. The quality of the material approximation at $\zr_e$ is measured as the distance $f_e(\zH^\circ_e,\zB^\circ_e)$ to the closest data point $(\zH^\times_{pq},\zB^\times_{pq})\in\mathcal{H}^\times$. The definition of $f_e(\zH^\circ_e,\zB^\circ_e)$ adopted here is inspired by the magnetic (co-)energy density (unit \si{\joule\per\meter\cubed}):
\begin{align}\label{eq:fe}
  \nonumber f_e(\zH^\circ_e,\zB^\circ_e) &=\min_{(\zH^\times_{pq},\zB^\times_{pq})\in\mathcal{H}^\times} 
  \Bigg[\frac{1}{2}(\zB^\circ_e-\zB^\times_{pq})^T\tilde{\nubf}_e(\zB^\circ_e-\zB^\times_{pq}) \\
  &\qquad+\frac{1}{2}(\zH^\circ_e-\zH^\times_{pq})^T\tilde{\mubf}_e(\zH^\circ_e-\zH^\times_{pq})\Bigg] \qcm
\end{align}
where $\tilde{\nubf}_e=\tilde{\mubf}^{-1}_e$ and $\tilde{\mubf}_e$ are user-defined reluctivity and permeability tensors employing a metric in phase space. The minimization problem \eqref{eq:fe} is a discrete one, i.e., for each material point $\zr_e$, the measurement point $(H^\times_{e,x},B^\times_{e,x})\in\mathcal{H}^\times_x$ which is closest to $(H^\circ_{e,x},B^\circ_{e,x})$ has to be found. Independently from that, the same is done for the $y$-direction. This operation is equivalent to partitioning both phase spaces as Voronoi diagrams (Fig.~\ref{fig:measurementdata}) \cite{Kirchdoerfer_2016aa}. For the entire field model, aligning the field solution with the data-based constitutive law is equivalent to minimizing the functional
\begin{align}\label{eq:F}
  F(\zH^\circ,\zB^\circ) &=\sum_{e\in\mathcal{N}\idx{elem}}
  f_e(\zH^\circ_e,\zB^\circ_e)S_e\ell_z \qpt
\end{align}
The overall optimization problem then reads
\begin{align}
  \nonumber\underset{\zH^\circ,\zB^\circ}{\text{minimize }} & F(\zH^\circ,\zB^\circ) \\
  \label{eq:opti}\text{subject to} & \left\{\begin{array}{rll}
    \zB^\circ &=\Cmat\afit \\
    \Cmat^T\Dmat_\Omega\zH^\circ &=\jfit
  \end{array}\right. \qpt
\end{align}
The discrete Maxwell equations are enforced by a Lagrange-multiplier approach, using the Lagrange multiplier $\etabf$. The data-driven field problem amounts to finding the stationary points of
\begin{align}\label{eq:augmented}
  \mathcal{F}(\zH^\circ,\afit,\etabf) &=F(\zH^\circ,\Cmat\afit)
  -\etabf^T(\jfit-\Cmat^T\Dmat_\Omega\zH^\circ) \qcm
\end{align}
as a function of $\zH^\circ$, $\afit$ and $\etabf$. The constrained minimization problem \eqref{eq:opti} is of a continuous nature, i.e., it distributes the magnetic field such that the total error on the data-based constitutive law is minimized while fulfilling the Maxwell's equations. At the linear-algebra side, this is carried out by solving a linear system of equations. The sub-minimization problems \eqref{eq:fe}, however, are of discrete nature and need an update between successive linear-system solves. Hence, the overall solver consists of an outer \emph{data-driven iteration} around an inner linear FE solve.

The initial solution for $\zH^\circ$ and $\zB^\circ$ is drawn randomly from the given data sets. For a temporary solution for $\zH^\circ$ and $\zB^\circ$ occurring during the data-driven iteration, the local minimizations \eqref{eq:fe} are carried out. The so-called \emph{active} measurement data, i.e., for each material point, $\zH^\times_{pq}$ and $\zB^\times_{pq}$ that are the closest to the solution for $\zH^\circ$ and $\zB^\circ$, are stored into the vectors $\zH^\times$ and $\zB^\times$. A variation of \eqref{eq:augmented} according to $\zH^\circ$ results in an explicit expression for $\zH^\circ$:
\begin{align}\label{eq:H}
  \zH^\circ &=\zH^\times+\Dmat_{\tilde{\nubf}}\Cmat\etabf \qpt
\end{align}
Variations with respect to the components of $\afit$ and $\etabf$ and inserting \eqref{eq:H} lead to the system of equations
\begin{align}\label{eq:system}
  \mtbb{\Knutld}{0}{0}{\Knutld}\mtba{\afit}{\etabf}
  &=\mtba{\Cmat^T\Dmat_\Omega\Dmat_{\tilde{\nubf}}\zB^\times}{\jfit-\Cmat^T\Dmat_\Omega\zH^\times} \qpt
\end{align}
Here, $\Knutld$ and $\Dmat_{\tilde{\nu}}$ are constructed as in \eqref{eq:Knu} and \eqref{eq:Dnu}, but using the user-defined $\tilde{\nubf}$ which defines the distance in phase space.

Essential boundary conditions on the magnetic vector potential need to be introduced. Thereby, possibly inhomogeneous Dirichlet data are applied to $\afit$, whereas homogeneous Dirichlet data are applied to the corresponding components of $\etabf$. Both blocks of the system can be solved independently. Because $\Kmat_{\tilde{\nubf}}$ is unaffected by the adaptation to the measurement data, its factorization can be carried out once and kept for all subsequent system solves. After solving \eqref{eq:system}, the new field approximation is retrieved by $\zB^\circ=\Cmat\afit$ and by \eqref{eq:H}. The procedure is repeated until the field solution does no longer change significantly.

It is easy to check that the formulation is consistent. In a situation with exact measurement data, the solution for $\zH\idx{sol}$ and $\zB\idx{sol}=\Cmat\afit\idx{sol}$ can be found among the measurement data, i.e., $\zH^\times=\zH\idx{sol}$ and $\zB^\times=\zB\idx{sol}$. Then, the first equation in \eqref{eq:system} is fulfilled exactly. Moreover, the right-hand side of the second equation is zero, which turns the Lagrange multiplier $\etabf$ to be zero as well, indicating that the minimum is found while satisfying the constraints exactly.

\section{Data-driven Field Solver:\\Inhomogeneous Material}

Many magnetic field problems feature an inhomogeneous material distribution, combining regions with exact material data and regions for which only measured data are available. Here, three approaches for dealing with this situation are distinguished.

\subsection{Approach~1: Exact material relation treated by the data-driven iteration}

The exact isotropic material characteristic $\zB=\mu\idx{ex}\zH$ can be represented in phase space by an infinite number of fictitious measurement points $(\zH^\times_{pq},\mu\idx{ex}\zH^\times_{pq})$, $\zH^\times_{pq}\in\mathbb{R}^2$, according to the exact material characteristic represented by the permeability $\mu\idx{ex}$. This observation allows to adapt the existing data-driven field solver in a minimal way. At first, the factors in the local minimization problem \eqref{eq:fe} in the exact regions are chosen as $\tilde{\mubf}=\diag{\mu\idx{ex}}$ and $\tilde{\nubf}=\diag{\nu\idx{ex}}$. Then, the solution of \eqref{eq:fe} is
\begin{subequations}
\begin{align}
  \zB^\times_e &=\frac{\zB^\circ_e+\mu\idx{ex}\zH^\circ_e}{2} \qscn
  \zH^\times_e &=\frac{\zB^\times_e}{\mu\idx{ex}} \qpt
\end{align}
\end{subequations}
The distance function then becomes
\begin{align}
  f\idxi{ex}{e}(\zH^\circ_e,\zB^\circ_e)
  &=\frac{1}{4}\nu\idx{ex}|\zB^\circ_e-\mu\idx{ex}\zH^\circ_e|^2 \qpt
\end{align}
The active (fictitious) measurement points can be updated in analogy to the data-driven materials, as part of the data-driven iteration steps. Moreover, the system~\eqref{eq:system} does not change, except for the fact that $\Dmat_{\nubf}$ is now inhomogeneous and contains the true reluctivities for the material points with exact data.


\subsection{Approach~2: Exact material relation minimized by the FE solver}

For the material points with exact data, the local minimization problems are of continuous nature and can therefore be shifted into the system of equations. As a result, the data-driven iteration only has to update the active measurement data for the data-driven material part. The computational domain $\Omega=\Omega\idx{a}\cup\Omega\idx{b}$ is partitioned in two non-overlapping domains, i.e., $\Omega\idx{a}$ containing the elements with exact material data and $\Omega\idx{b}$ containing the element with measurement data. The set of triangles is partitioned accordingly, i.e., $\mathcal{N}\idx{elem}=\mathcal{N}\idxi{elem}{a}\cup\mathcal{N}\idxi{elem}{b}$. Furthermore, also the operators $\Cmat$, $\Dmat_{\tilde{\nubf}}$ and $\Dmat_\Omega$ can be partitioned accordingly, i.e., $\Cmat^T=\mtab{\Cmat\idx{a}^T}{\Cmat\idx{b}^T}$, $\Dmat_{\tilde{\nubf}}=\diag{\mtab{\Dmat_{\nu\idx{a}}}{\Dmat_{\tilde{\nubf}\idx{b}}}}$ and $\Dmat_{\Omega}=\diag{\mtab{\Dmat_{\Omega\idx{a}}}{\Dmat_{\Omega\idx{b}}}}$.
The functional then reads $F(\zH^\circ,\zB^\circ)=F\idx{a}(\zH^\circ\idx{a},\zB^\circ\idx{a})+F\idx{b}(\zH^\circ\idx{b},\zB^\circ\idx{b})$, where the indices $\cdot\idx{a}$ and $\cdot\idx{b}$ restrict the vectors to the field data for $e\in\mathcal{N}\idxx{elem}{a}$ and $e\in\mathcal{N}\idxx{elem}{b}$, respectively, and \begin{subequations}
\begin{align}
  F\idx{a}(\zH^\circ\idx{a},\zB^\circ\idx{a}) &=\sum_{e\in\mathcal{N}\idxi{elem}{a}}f\idxi{ex}{e}(\zH^\circ_e,\zB^\circ_e)S_e\ell_z \qscn
  F\idx{b}(\zH^\circ\idx{b},\zB^\circ\idx{b}) &=\sum_{e\in\mathcal{N}\idxi{elem}{b}}f_e(\zH^\circ_e,\zB^\circ_e)S_e\ell_z \qpt
\end{align}
\end{subequations}

The functional is extended to include Amp\`ere's law, similarly to \eqref{eq:augmented}. Then, the variations according to $\zH^\circ\idx{a}$ and $\zH^\circ\idx{b}$ provide the relations
\begin{subequations}
\begin{align}
  \label{eq:Ha}\zH^\circ\idx{a} &=\Dmat_{\nu\idx{a}}\Cmat\idx{a}(\afit-\etabf) \qscn
  \label{eq:Hb}\zH^\circ\idx{b} &=\zH^\times\idx{b} +\Dmat_{\tilde{\nubf}\idx{b}}\Cmat\idx{b}\etabf \qcm
\end{align}
\end{subequations}
respectively. By expressing the variations according to $\afit$ and $\etabf$ and inserting \eqref{eq:Ha} and \eqref{eq:Hb}, one finds the system of equations
\begin{align}
\mtbb{\Kmat_{\tilde{\nubf}\idx{b}}}{-\Kmat_{\nu\idx{a}}}{\Kmat_{\nu\idx{a}}}{\Kmat_{\tilde{\nubf}\idx{b}}+2\Kmat_{\nu\idx{a}}} \mtba{\afit}{\etabf} &=\mtba{\Cmat\idx{b}^T\Dmat_{\Omega\idx{b}}\Dmat_{\tilde{\nubf}\idx{b}}\zB^\times\idx{b}}{\jfit-\Cmat\idx{b}^T\Dmat_{\Omega\idx{b}}\zH^\times\idx{b}} \qcm
\end{align}
where
\begin{subequations}
  \begin{align}
  \Kmat_{\tilde{\nubf}\idx{b}}
  &=\Cmat\idx{b}^T\Dmat_{\Omega\idx{b}}\Dmat_{\tilde{\nubf}\idx{b}}\Cmat\idx{b} \qscn
  \Kmat_{\nu\idx{a}}
  &=\Cmat\idx{a}^T\Dmat_{\Omega\idx{a}}\Dmat_{\nu\idx{a}}\Cmat\idx{a} \qpt
  \end{align}
\end{subequations}
Both subsystems can no longer be solved independently. Moreover, $\Kmat_{\tilde{\nubf}\idx{b}}$ and $\Kmat_{\nu\idx{a}}$ are not regular and, hence, cannot be factorized. However, as long as the exact material data are linear, the block system as a whole remains constant during the data-driven iteration and can be factorized once in advance and then be applied repeatedly. The system solution is more costly than for the first approach.

\subsection{Approach~3: Exact material relation enforced by the FE solver}

Approach~2 resolves the exact material relation by minimizing $F_a(\zH^\circ\idx{a},\zB^\circ\idx{a})$ together with the remaining part of the functional and in combination with the constraints. Hence, the relation will not be fulfilled exactly, as is also clear from the shape of \eqref{eq:Ha} where the Lagrange multipliers $\etabf$ may not be zero. One can still go one step further and enforce the exact material relation. Then, one solves
\begin{align}\label{eq:augmented3}
  \nonumber \lefteqn{\min_{\zH^\circ\idx{b},\afit,\etabf} F\idx{b}(\zH^\circ\idx{b},\Cmat\idx{b}\afit)} &
  \\ &-\etabf^T(\jfit-\Cmat^T\idx{a}\Dmat_{\Omega\idx{a}}\Dmat_{\nu\idx{a}}\Cmat\idx{a}\afit
    -\Cmat^T\idx{b}\Dmat_{\Omega\idx{b}}\zH^\circ\idx{b}) \qpt
\end{align}
After a bit of calculus, one finds
\begin{subequations}
  \begin{align}
  \zH^\circ\idx{a} &=\Dmat_{\nu\idx{a}}\Cmat\idx{a}\afit              \qscn
  \zH^\circ\idx{b} &=\zH^\times\idx{b} +\Dmat_{\tilde{\nubf}\idx{b}}\Cmat\idx{b}\etabf \qcm
  \end{align}
\end{subequations}
as the update for the magnetic field strengths and
\begin{align}
  \mtbb{\Kmat_{\tilde{\nubf}\idx{b}}}{-\Kmat_{\nu\idx{a}}}{\Kmat_{\nu\idx{a}}}{\Kmat_{\tilde{\nubf}\idx{b}}} \mtba{\afit}{\etabf} &=\mtba{\Cmat\idx{b}^T\Dmat_{\Omega\idx{b}}\Dmat_{\tilde{\nubf}\idx{b}}\zB^\times\idx{b}}{\jfit-\Cmat\idx{b}^T\Dmat_{\Omega\idx{b}}\zH^\times\idx{b}} \qcm
\end{align}
as the system to be solved.

\section{Example: Quadrupole Magnet}

\begin{figure}[t]
  \centering
  \scalebox{1}[-1]{\includegraphics[width=6cm]{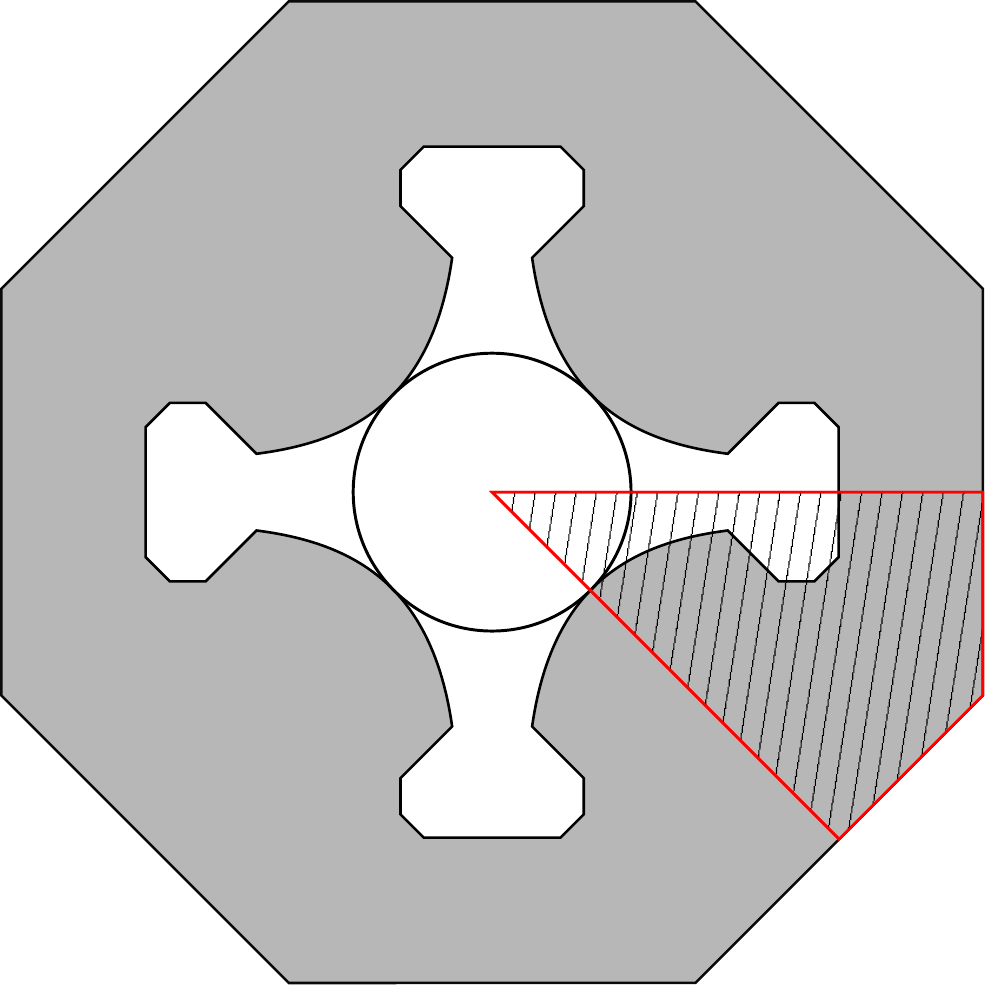}}
  \caption{Quadrupole magnet: cross section, modeled part.}
  \label{fig:quadgeometry}
\end{figure}

The anisotropic data-driven magnetostatic field solver is applied to the 2D model of a quadrupole accelerator magnet (Fig.~\ref{fig:quadgeometry}). The symmetry of the geometry and the excitations allow to consider only one eighth of the magnet's circumference. The iron has a relative permeability of $300$ in the $y$-direction. The permeability in the $x$-direction is given by the material curve of Fig.~\ref{fig:measurementdata}. The permeability of the yoke is high enough to neglect the magnetic flux leaving the outer boundary of the yoke. This assumption allows to apply electric boundary conditions to the yoke's outer boundary. The simulation aims at the magnetic flux density in the magnet's aperture for the nominal current in the windings. For that purpose, a nonlinear magnetostatic FE simulation is appropriate. The reference solution is obtained with this material curve and for a 2D FE mesh with $N\idx{node}=3277$ nodes and $N\idx{elem}=6344$ elements.

\begin{figure*}[t]
  \centering
  \begin{subfigure}[t]{0.30\textwidth}
    \includegraphics[width=\textwidth]{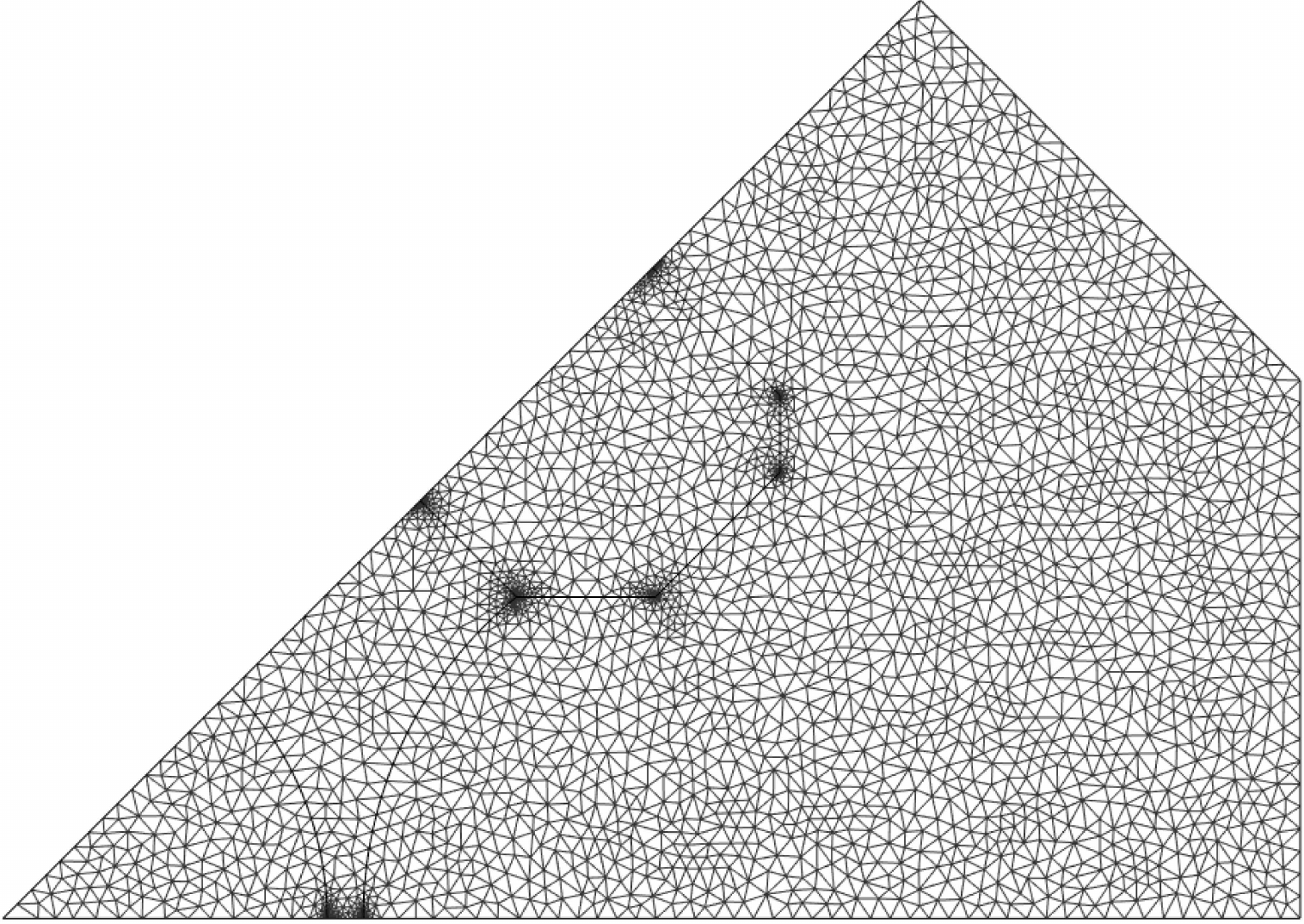}
    \caption{Mesh}
    \label{fig_quad_mesh}
  \end{subfigure}
  \begin{subfigure}[t]{0.30\textwidth}
    \includegraphics[width=\textwidth]{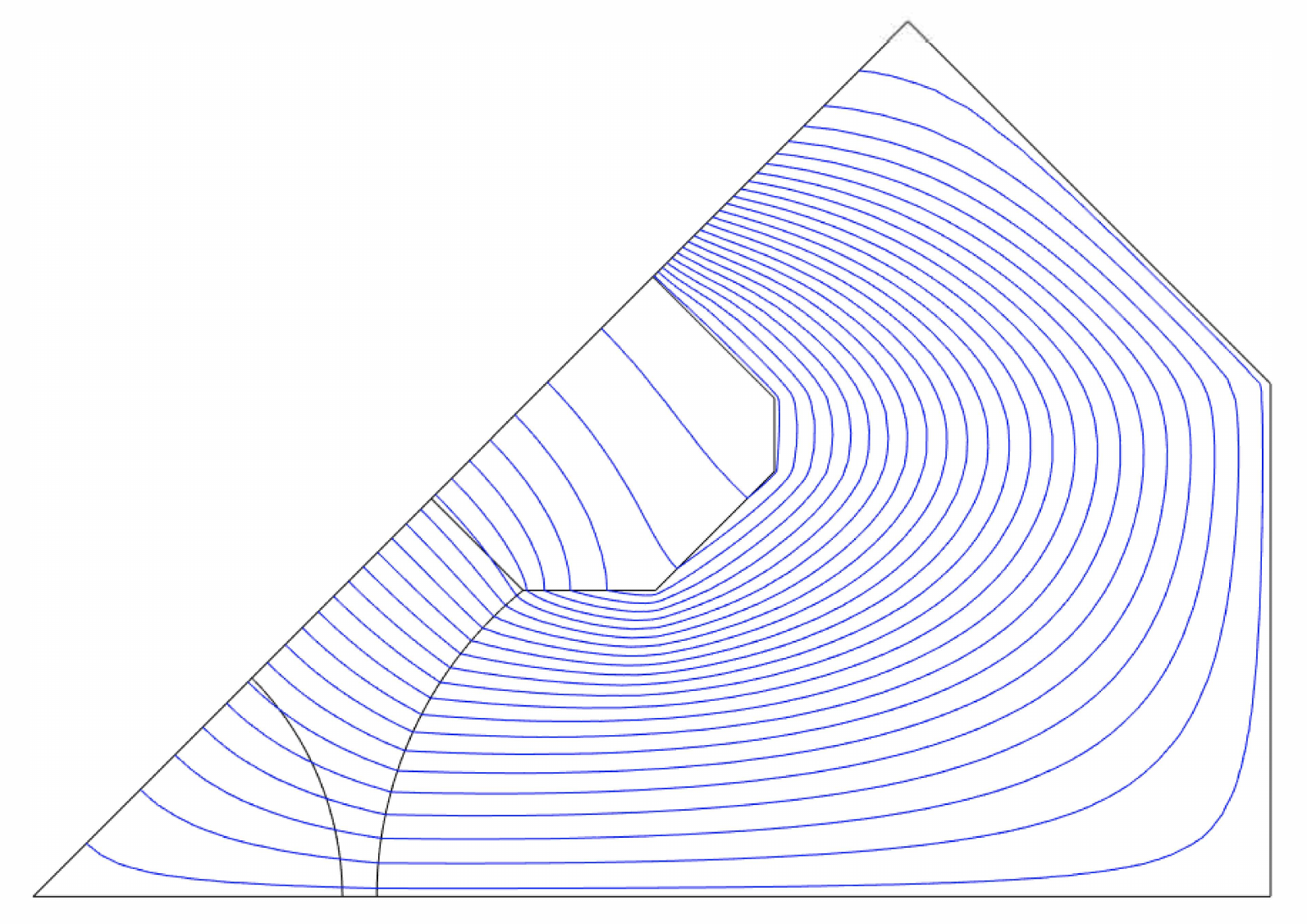}
    \caption{Magnetic flux lines}
    \label{fig_quad_equipots}
  \end{subfigure}
  \begin{subfigure}[t]{0.30\textwidth}
    \includegraphics[width=\textwidth]{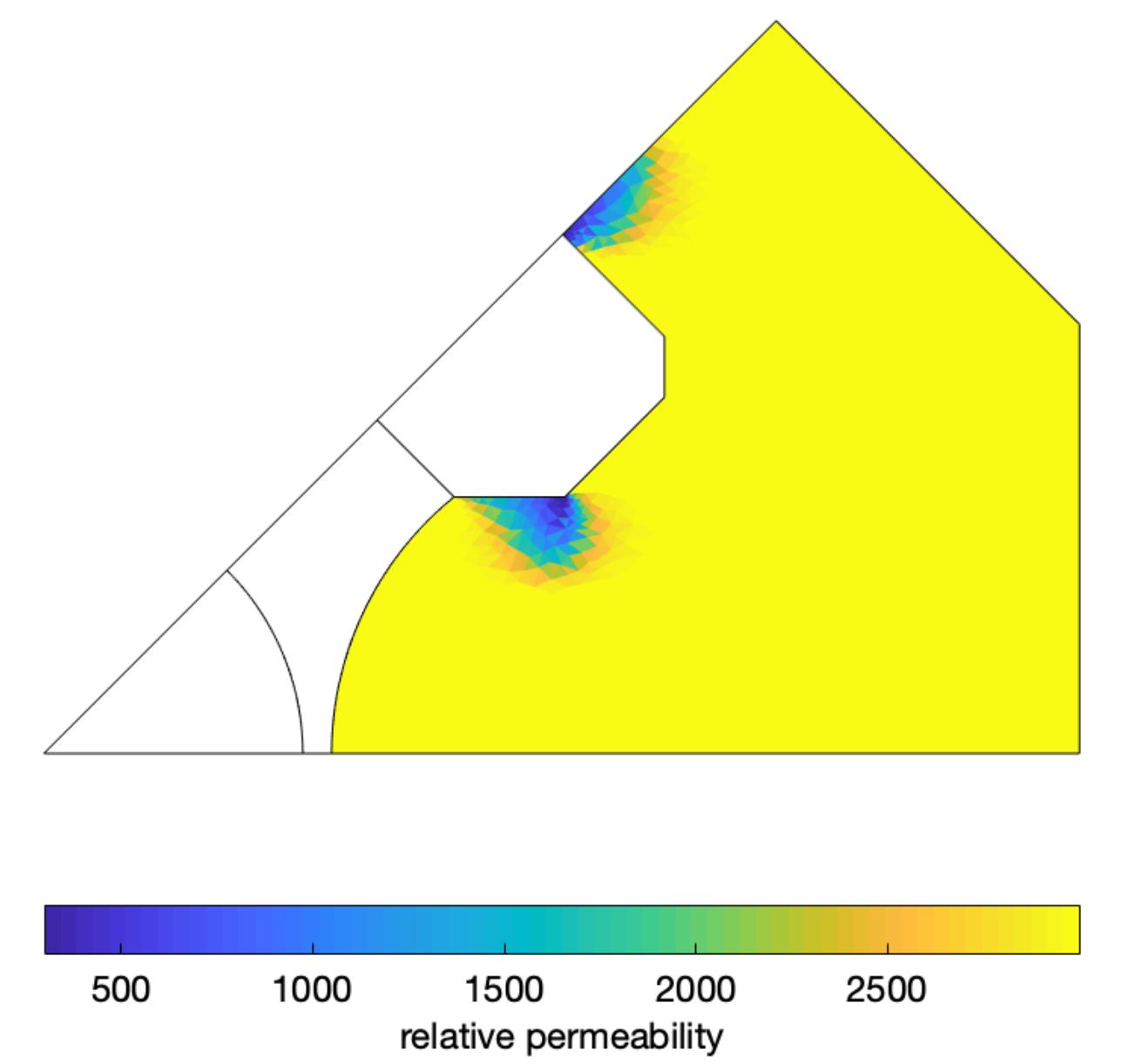}
    \caption{Relative permeability in iron}
    \label{fig_quad_relperm}
  \end{subfigure}
  \caption{Quadrupole magnet.}
  \label{fig:quad}
\end{figure*}
\begin{figure*}[t]
  \centering
  \begin{subfigure}[b]{0.45\textwidth}
    \includegraphics[width=7cm]{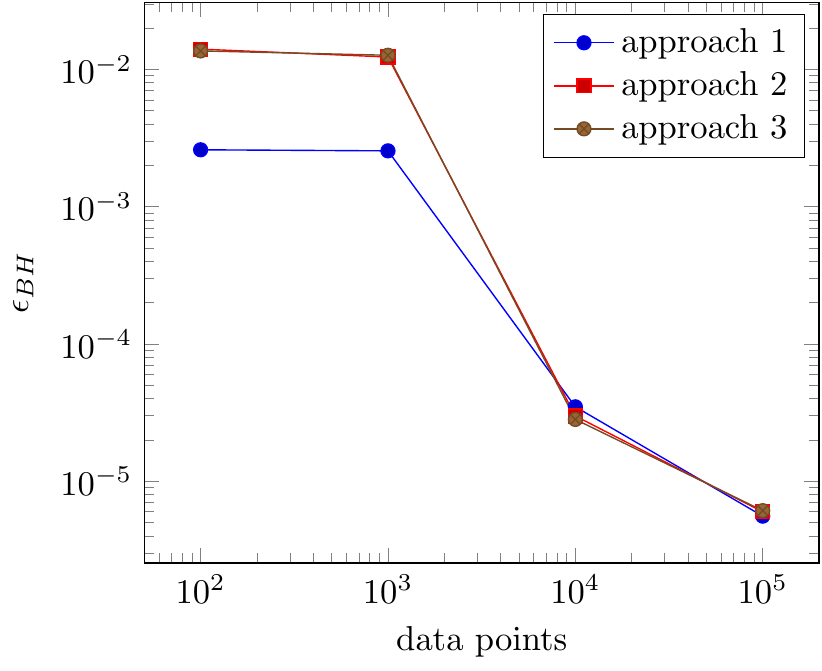}
    \caption{material modeling error}
    \label{fig:quadconvmatmoderr}
  \end{subfigure}  
  \begin{subfigure}[b]{0.45\textwidth}
    \includegraphics[width=7cm]{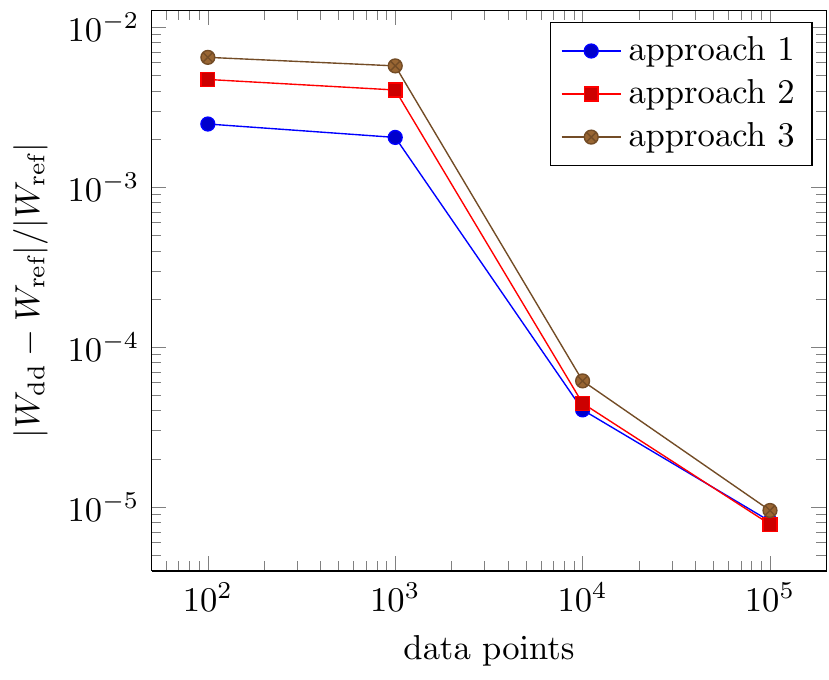}
    \caption{relative error on the magnetic energy}
  \end{subfigure}
  \caption{Convergence of the data-driven procedure with respect to the number of data points.}
  \label{fig:quadconvergence}
\end{figure*}
\begin{figure*}[t]
  \begin{subfigure}[b]{0.33\textwidth}
    \includegraphics[width=5.5cm]{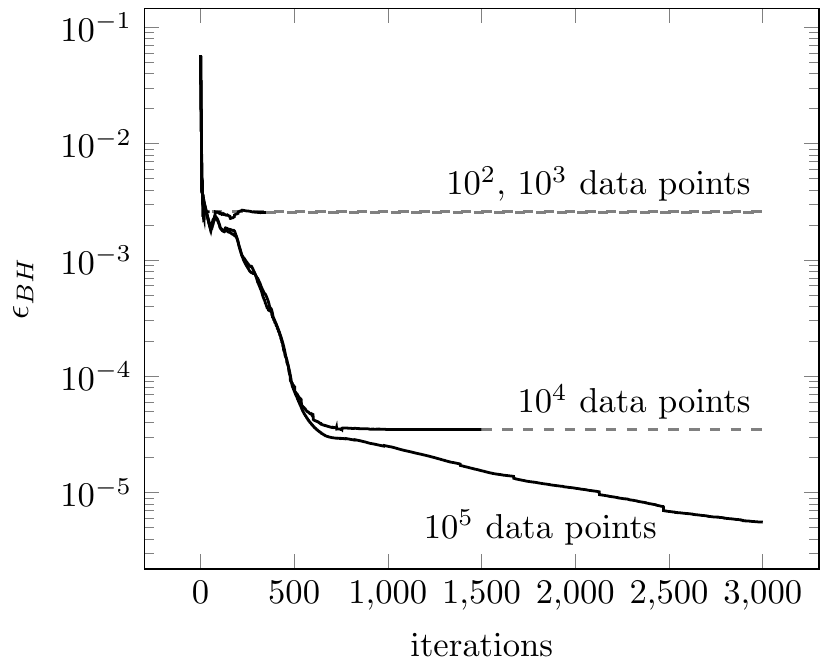}
    \caption{Approach 1}
  \end{subfigure}
  \begin{subfigure}[b]{0.33\textwidth}
    \includegraphics[width=5.5cm]{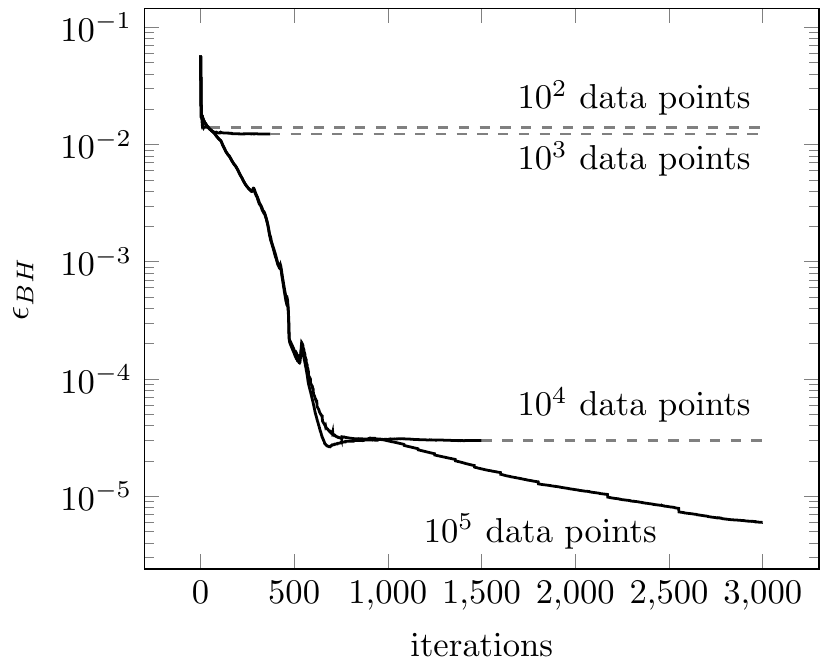}
    \caption{Approach 2}
  \end{subfigure}
  \begin{subfigure}[b]{0.33\textwidth}
    \includegraphics[width=5.5cm]{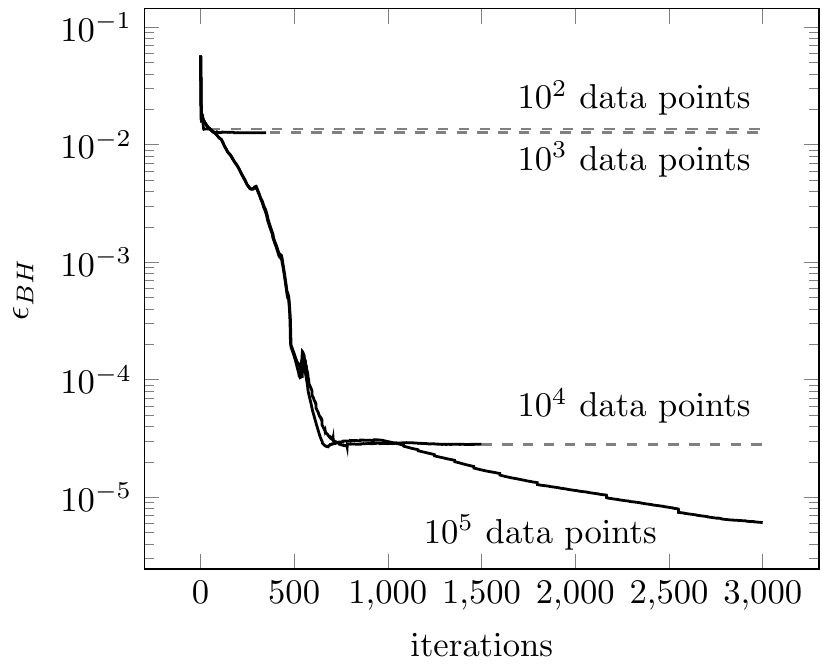}
    \caption{Approach 3}
  \end{subfigure}
  \caption{Convergence of the material modeling error according to the number of data-driven iterations.}
  \label{fig:quaditerations}
\end{figure*}

The material curve is replaced by a set of $N\idx{data}\in\{10^2,10^3,10^4,10^5\}$ data points, equidistantly distributed along the curve. Furthermore, the data are assumed to be noise-free. The finite set of data points cause a \emph{material modeling error} characterized by
\begin{align}\label{eq:matmoderr}
\epsilon_{BH} &=\sqrt{\frac{1}{N\idx{elem}}\sum_{e=1}^{N\idx{elem}} S_e\ell_z\left(\epsilon_{H_x,e}^2+\epsilon_{H_y,e}^2+\epsilon_{B_x,e}^2+\epsilon_{B_y,e}^2\right)}\qscn
  \epsilon_{X_d,e}^2 &=\frac{\left|X_{e,d}-X^\S_{e,d}\right|^2}{\left|X^\S_{e,d}\right|^2} \qcm
\end{align}
where $X^\S_{e,d}$ denotes the $d$-component at element~$e$ of the field value $X$ of the reference solution. Additionally, the relative error between the magnetic energy $W\idx{dd}$ solved by the data-driven approach and the magnetic energy $W\idx{ref}$ of the reference solution is monitored. The calculations are carried out for each of the three above developed approaches. The solution is shown in Fig.~\ref{fig:quad}.

The convergence of the data-driven model towards the reference model is shown in Fig.~\ref{fig:quadconvergence}. For each of the three approaches, more precise material data decrease both errors similarly. The convergence of the material modeling error according to the data-driven iteration is depicted in Fig.~\ref{fig:quaditerations}. Obviously, the convergence stagnates at the material modelling error related to the used data set. Moreover, all three approaches perform equally well. The numerical tests indicate that a moderate accuracy of, e.g., $10^{-2}$ can be attained with a sufficient amount of data points, here, $10^2$, for all three apporaches. However, the first approach needs $9$~iterations to reach the desired accuracy, whereas the approaches~2 and~3 need only $2$, respectively $3$ iterations. For a moderate number of data points, approach~1 performs slightly better than the other approaches. However, this comes at the cost of more iterations. In the case of a dense data set, all three approaches perform well and similar results are obtained. The number of iterations is acceptable, considering the fact that the Newton approach applied for obtaining the reference solution requires $20$~iterations for attaining a relative error of $10^{-6}$ of the Newton iteration. When a better accuracy is needed, more data points are mandatory. Then, however, the number of needed data-driven iterations increases dramatically (Fig.~\ref{fig:quaditerations}).

\section{Conclusion}
It is possible to build a \emph{data-driven FE solver} exploiting a measured material data set directly, as an alternative to standard FE simulation which relies upon a  material curve deduced by regression. To that purpose, the standard nonlinear FE procedure is replaced by an outer data-driven iteration around a modified linear FE solver. The data-driven FE solver is further developed to deal with anisotropic material data and for cases where measured and exact material data coexist. For the 2D model of a quadrupole magnet, less than $10$ data-driven iterations were needed to achieve convergence when a moderate accuracy is acceptable. The improved approaches proposed in the paper for models which combine exactly-known and data-driven material information, show a faster convergence when only a moderate number of data points is at hand.

\appendices
\section*{Acknowledgment}
This work has been supported by the Deutsche Forschungsgemeinschaft (DFG), Research Training Group 2128 ''Accelerator Science and Technology for Energy Recovery Linacs''. The authors thank Laura A. M. D'Angelo and Markus Borkowski for their help in maintaining the simulation software.

\bibliographystyle{abbrv}
\bibliography{tmag_degersem_bib}

\end{document}